\documentclass[final,5p,times,twocolumn]{elsarticle}

\usepackage{amssymb,amsmath}

\journal{\;}

\begin{document}

\begin{frontmatter}



\title{Power law susceptibility function for the analysis of anomalous spectral response}


\author[inst1,camr,fiu]{Anis Allagui\corref{cor1}}

\affiliation[inst1]{organization={Dept. of Sustainable and Renewable Energy Engineering},
            addressline={University of Sharjah}, 
            city={Sharjah},
            postcode={27272}, 
            state={Sharjah},
            country={United Arab Emirates}}

\affiliation[camr]{organization={Center for Advanced Materials Research, Research Institute of Sciences and Engineering},
            addressline={University of Sharjah}, 
            city={Sharjah},
            postcode={27272}, 
            state={Sharjah},
            country={United Arab Emirates}}

\affiliation[fiu]{organization={Dept. of Electrical and Computer Engineering},
            addressline={Florida International University}, 
            city={Miami},
            postcode={33174}, 
            state={FL},
            country={United States}}
                        
\ead{aallagui@sharjah.ac.ae}
\cortext[cor1]{Corresponding author}

\author[inst3]{Enrique H. Balaguera}
\affiliation[inst3]{organization={Escuela Superior de Ciencias Experimentales y Tecnologia, Universidad Rey Juan Carlos},
            addressline={C/ Tulip{a}n, s/n}, 
            city={Mostoles},
           postcode={28933},
            state={Madrid},
            country={Spain}}

\begin{abstract}

The  extensions of the classical Debye model of susceptibility of dielectric materials to the well-known Cole-Cole,  Davidson-Cole, or the Havriliak-Negami models is done by introducing   non-integer power parameters to the frequency-domain function. This is very often necessary in order to account for anomalous deviations of the experimental data from the ideal case. The corresponding time-domain descriptions expressed in terms of the relaxation or response functions are in the form of  first-order differential equations for the case of Debye model, but involves relatively complex integro-differential operators for the modified ones. 
In this work, we study the extension of the time-domain  kinetic equation describing the Debye polarization function to include two extra degrees of freedom; one to transform the first-order time derivative of the polarization function to the Caputo fractional-order time derivative and another to change the linear term to a power term. From an electrical perspective, it results in a constant-phase element with two fractional parameters.

\end{abstract}




\end{frontmatter}

\section{Introduction}

Experimental dielectric spectroscopy data are commonly interpreted using the phenomenological relaxation models of Cole-Cole \cite{cole1941dispersion}, Davidson-Cole \cite{davidson1951dielectric}, or Havriliak-Negami \cite{havriliak1966complex}. These well-known equations can be viewed as direct extensions of the standard Debye model of dielectrics by inserting one or two real powers into the frequency-domain function. In the time domain, whether it is for the case of decrease in the absorption current by the dielectric material after the application of a DC voltage (Curie-von Schweidler law \cite{curie1888recherches,v1907studien}), or the potential decay after a surface charge deposit, the response of complex dielectric systems often follows power law profiles\;\cite{jonscher1977universal}. The Debye processes indeed lead to exponential relaxations with a single characteristic time constant, which is in most of the cases not satisfactory enough for fitting experimental data\;\cite{garrappa2016models}. Electrically speaking, this classical model is associated with the charging/discharging of ideal capacitors. In contrast, Cole-Cole, Davidson-Cole, and Havriliak-Negami patterns correlate to generalized electrical elements, such as the famous constant phase element (CPE) \cite{lasia2022origin}, and the ideal/generalized Gerischer impedance \cite{gerischer1951wechselstrompolarisation}, universally used in the electrical characterization of complex materials via impedance spectroscopy. Another approach towards extending Debye processes is based on the distribution function of relaxation times\;\cite{ciucci2015analysis}, which considers the system to be a discrete or a continuous distribution of Debye models with relaxation times derived from a function that characterizes the actual system under consideration\;\cite{dfrt}. This approach has been indeed proven to have a great potential in analyzing spectral data of different nature\;\cite{boukamp2017analysis, DIERICKX2020136764, song2023non}.

In this work, we seek to analyze an alternative generalization of the Debye model considering the time-domain differential equation governing the evolution of the dielectric system as the starting point. More specifically, we propose a nonlinear fractional-order differential equation for the kinetics of the polarization function wherein its rate, taken in a general sense as the Caputo time fractional derivative with respect to time, is equal to a power law function of the same variable. For the standard case, one has a first-order time derivative of the polarization function equal to a linear function of the same. Our approach leads to Eq.\;\ref{eqab} below which has an analytical power law solution. From the time-domain polarization function, we obtain the frequency-domain susceptibility function of the system for which we analyze its salient features   from the perspective of electrical network analysis and synthesis.

\section{Theory}

\subsection{Basic definitions}
The complex, frequency-dependent susceptibility function of dielectrics is given in its normalized form by \cite{garrappa2016models}:
\begin{equation}
\tilde{\chi}(s) = \frac{\tilde{\epsilon}(s)-\epsilon_{\infty}}{\epsilon_0 - \epsilon_{\infty}} 
\label{eq1}
\end{equation}
where $s=j \omega$, $\omega=2\pi f$ is the complex frequency, 
and $\tilde{\epsilon}(s)$, $\epsilon_0$ and $\epsilon_{\infty}$ are the complex permittivity of the dielectric and its low and high-frequency limits, respectively. The pulse-response function of the system denoted by $\phi(t)$, i.e. the normalized transient current which flows when the steady macroscopic electric field is removed,  is obtained by inverse Laplace transform (LT) of the susceptibility function as \cite{garrappa2016models}:
\begin{equation}
    \phi(t) = \mathcal{L}^{-1} \left[ \tilde{\chi}(s) ; t \right] 
    \label{eq1}
\end{equation}
We recall that the LT of a real function $f(t)\;(t\in\mathbb{R}^+)$  is defined by $\mathcal{L}[f(t);s] = F(s)= \int_0^{\infty} e^{-st} f(t) dt,\;s\in\mathbb{C}$, and the inverse LT of $F(s)$  is defined by $\mathcal{L}^{-1}[F(s);t] = f(t) = ({2\pi i})^{-1} \int_{\gamma-i\infty}^{\gamma+i\infty} e^{st} F(s) ds,\;\mathcal{R}e(s)=\gamma $.  
The relaxation or decay function   of the polarization ($\Psi(t)$) of the system when a steady  field is removed  is derived from \cite{garrappa2016models}:
\begin{equation}
    \Psi(t) = 1- \mathcal{L}^{-1} \left[ s^{-1} \tilde{\chi}(s);t \right]
    \label{eq:Psi}  
\end{equation} 
These two   functions are related as follows:
\begin{equation}
\phi(t) = -\frac{\mathrm{d}\Psi(t)}{ \mathrm{d} t}
    \label{eq3}
\end{equation}
They are both causal functions of time which means they vanish for $t<0$ \cite{garrappa2016models}. In the Laplace domain, we have from Eq.\;\ref{eq3}  with the use of Eq.\;\ref{eq1}, the following relation between the relaxation function and the pulse-response function:
\begin{equation}
\mathcal{L}  \left[  {\Psi}(t) ; s \right] 
= \frac{1-  \mathcal{L}  \left[  {\phi}(t) ; s \right] }{s}
\end{equation}
In what follows, we show a few examples of these functions for a few of the well-known classical theories.

\subsection{Classical models}

It is clear from Eq.\;\ref{eq:Psi} for the polarization function that for the case of a simple exponential decay  of the (normalized) form 
\begin{equation}
\Psi_D(t) = e^{-t/\tau_D}
\end{equation}
 where $\tau_D$ is a characteristic relaxation time constant, one obtains the classical single-pole    Debye model for the susceptibility function as:
\begin{equation}
\tilde{\chi}_D(s)  
=   \frac{1}{1+s \tau_D}
\end{equation}
The function $\Psi_D(t)$ also satisfies the kinetic equation (initial value problem):
\begin{equation}
\frac{\mathrm{d}\Psi_D(t)}{\mathrm{d}t}+ \tau_D^{-1}{{\Psi_D(t)}}=0, \quad \Psi_D(0)=1 
\label{eq:psid}
\end{equation}
Then the response function $\phi_D(t)$ is given by:
\begin{equation}
    \phi_D(t)  = {\tau_D^{-1}} e^{-t/\tau_D}
\end{equation}
which itself satisfies the differential equation:
\begin{equation}
\frac{\mathrm{d}\phi_D(t)}{\mathrm{d}t}+ \tau_D^{-1}{{\phi_D(t)}}=0, \quad \phi_D(0)= \tau_D^{-1} 
\label{eqrho}
\end{equation}

But as revealed by a large amount of experimental results,  
the physical relaxation phenomena in materials which kinetics is influenced by disorder, complexity, or randomness exhibit a broad distribution of relaxation times and thus cannot be described with the  simple theory of Debye, but rather with  power-law  models. The origin of this can be attributed in the general sense to many-body interactions within the material  \cite{baker2000generalized}.
The   Cole-Cole \cite{cole1941dispersion}, Davidson-Cole \cite{davidson1951dielectric} and Havriliak-Negami \cite{havriliak1966complex} dispersion models extend the classical Debye theory by introducing one or two extra fractional exponents directly in the expression of the  susceptibility function \cite{hilfer2002analytical, goychuk2007anomalous, cpe}. For instance, the Cole-Cole equation in the Laplace domain is given by  \cite{cole1941dispersion}: 
\begin{equation}
\tilde{\chi}_{\alpha}(s) = \frac{1}{1 + (s\tau_{\alpha}) ^{\alpha}} \quad (0<\alpha \leqslant 1)
\label{eqCC}
\end{equation}
where $\tau_{\alpha}$ is a reference relaxation time constant, and from Euler's   formula we recall that $i^{\alpha}= \cos(\alpha \pi/2) + i \sin (\alpha \pi/2)$. 
Eq.\;\ref{eqCC} readily simplifies to Debye's model when the power coefficient $\alpha=1$. 
 The corresponding  time-domain   relaxation  function is given (in normalized form) by: \cite{garrappa2016models,10.1149/1945-7111/ac621e}:
\begin{equation}
\Psi_{\alpha}(t) = E_{\alpha,1}^1\left[ -(t/\tau_{\alpha})^{\alpha} \right]
\label{eq:Psialpha}
 \end{equation}  
 when the initial condition is taken to be  $\Psi_{\alpha}(0)=1$. Here
 \begin{equation}
{E}_{a,b}^{c} ( z ) := \sum\limits_{k=0}^{\infty} \frac{(c)_k}{\Gamma(a k + b)} \frac{z^k}{k!} \quad (a,b, c \in \mathbb{C}, \mathrm{Re}({a})>0)
\label{eqML}
\end{equation}
 (with $\Gamma(x)$ is the gamma function, $(c)_k = c(c+1)\ldots(c+k-1) =\Gamma(c+k)/\Gamma(c)$ being the Pochhammer symbol) is the three-parameter or generalized Mittag-Leffler  function. 
Eq.\;\ref{eq:Psialpha} is   the solution to the fractional differential equation \cite{garrappa2016models}:
\begin{equation}
 {{}_0^CD_t^{\alpha}  \Psi_{\alpha}(t) + \tau_{\alpha}^{-1}{\Psi_{\alpha}(t) } }=0
 \label{eqFKCC}
\end{equation}
 where the operator ${}^C_0 D_t^{\alpha} f(t)$ represents the  Caputo fractional derivative of $f(t)$: 
 \begin{equation}
{}^C_0 D_t^{\alpha} f(t) = \frac{1}{\Gamma(m-\alpha)} \int_0^t (t-\tau)^{m-\alpha-1} f^{(m)}(\tau) d\tau
\end{equation}
where $m\in \mathbb{N}$, $m-1< \alpha \leqslant m$ (or $m =\lceil{\alpha}\rceil $), and $f^{(m)}(t)=d^m f(t)/dt^m$. 
 Its Laplace transform is given by:
\begin{equation}
\mathcal{L}\left[^C_0D_t^{\alpha} f(t); s \right]= s^{\alpha} F(s) - \sum\limits_{k=0}^{m-1} s^{\alpha-k-1} f^{(k)}(0^+)
\end{equation} 
We can also write
 \cite{khamzin2014justification}:
\begin{equation}
\frac{\mathrm{d}\Psi_{\alpha}(t)}{\mathrm{d}t} +\tau_{\alpha}^{-\alpha} \, _0D_t^{1-\alpha}\Psi_{\alpha}(t)  =0
\label{eq217}
\end{equation}
 with $\Psi_{\alpha}(0)=1$, and  
 $_0D_t^{1-\alpha} f(t) = (\mathrm{d}/\mathrm{d}t)_0D_t^{-\alpha} f(t)$ denotes  the fractional derivate in the Riemann-Liouville sense  with:
  \begin{equation}
 _0D_t^{-\alpha} f(t)= _0I_t^{\alpha} f(t) =  \frac{1}{\Gamma(\alpha)} \int_0^t    {(t-\tau)^{\alpha-1}}   f(\tau)   d\tau
 \label{eqRL1}
\end{equation}
being the right-sided  Riemann-Liouville fractional integral of order $\alpha$. 
These fractional differentiation and integration operators involve a convolution integral,  which can be regarded as a consequence of memory effects \cite{ryabov2002novel}.

Similarly,    the   Davidson-Cole susceptibility model with one extra, non-integer parameter $\gamma$,  and its associated relaxation function with its governing constitutive equation are given, respectively, by      \cite{davidson1951dielectric, rosa2015relaxation, hilfer2002analytical, garrappa2016models}:
\begin{align}
& \tilde{\chi}_{\gamma}(s) = \frac{1}{(1+ s \tau_{\gamma})^{\gamma}} \quad  (0<\gamma \leqslant 1) \\
& \Psi_{\gamma}(t) =1- (t/\tau_{\gamma})^{\gamma} E_{1,\gamma+1}^{\gamma} (-t/\tau_{\gamma}) \\
& \frac{\mathrm{d}\Psi_{\gamma}(t)}{\mathrm{d}t} +
\tau_{\gamma}^{-\gamma}
\frac{\mathrm{d}}{\mathrm{d}t}
\left\{ e^{-t/\tau_{\gamma}} \mathrm{\mathbf{E}}^1_{\gamma,\gamma,\tau_{\gamma}^{-\gamma},0+} e^{\tau/\tau_{\gamma}} \Psi_{\gamma}(t) \right\}
 = 0
\label{eqDC}
\end{align}
 ($\tau_{\gamma}$ is a reference   time constant), and for  the more general case of Havriliak-Negami  model (with the two extra parameters $\alpha$ and $\gamma$)   \cite{havriliak1966complex} we have the following results \cite{khamzin2014justification, rosa2015relaxation, hilfer2002analytical, garrappa2016models}:
\begin{align}
&{\Psi_{H}(s) = \frac{1}{\left(1+ (s \tau_H)^{\alpha}\right)^{\gamma}}} \quad ( 0< \alpha, \gamma  \leqslant 1) \\
& \Psi_{H}(t)= 1- (t/\tau_H)^{\alpha \gamma} E_{\alpha,\alpha\gamma+1}^{\gamma}\left[- (t/\tau_H)^{\alpha} \right] \\
& \frac{\mathrm{d}\Psi_{H}(t)}{\mathrm{d}t}
 +
\sum\limits_{k=0}^{\infty} \tau_H^{-\alpha\gamma(k+1)} \frac{\mathrm{d}}{\mathrm{d}t}  \mathrm{\mathbf{E}}^{\gamma(k+1)}_{\alpha,\alpha\gamma(k+1),-\tau_H^{-\alpha},0+} \Psi_{H}(t)
 = 0
\label{eqH}
\end{align}
 where \begin{equation}
\mathrm{\mathbf{E}}_{\rho,\mu,\omega,a+}^{\gamma}  f(t) 
= \int_a^t (t-\tau)^{\mu-1} E_{\rho,\mu}^{\gamma} [\omega(t-\tau)^{\rho}] f(\tau) d\tau 
\end{equation}
is the Kilbas-Saigo-Saxena integral operator \cite{kilbas2004generalized} ($\tau_{H}$ is another  specific   time constant).
 It is clear that the Davidson-Cole   model simplifies to the Debye model for $\gamma=1$, and the Havriliak-Negami  model simplifies to the Cole-Cole model for $\gamma=1$, to the Davidson-Cole   model for $\alpha=1$, and to the Debye model for both $\alpha=1$ and $\gamma=1$.

\subsection{The proposed model}


We study an alternative extension of the Cole-Cole model, which  is again itself an extension of the Debye model, by having Eq.\;\ref{eqFKCC} describing the relaxation process of the system according to  Cole-Cole to be    instead   the following two-parameter homogeneous fractional nonlinear differential equation:
 \begin{equation}
    ^C_0D_t^{\alpha} \Psi(t) - q_0 \Psi(t)^{\beta} =0; \;\; \bf{\Psi(0)=0}
    \label{eqab}
\end{equation}
Here  again for the Caputo fractional derivative $\lceil{\alpha}\rceil =m$,  but  specifically for our case $m=1$, $\beta \in \mathbb{R}$ ($\beta \neq 0,1$) and $q_0 \in \mathbb{R}$ ($q_0 \neq 0$). 
We do not attempt to suggest any physical interpretation for the parameter $\beta$ as this time. The rest of the manuscript is focused solely on the derivation and exploration of the mathematical results that can be deducted from Eq.\;\ref{eqab}. 
Otherwise it is clear from the proposed Eq.\;\ref{eqab} that for the case of both $\alpha=1$ and $\beta=1$, one recovers the evolution differential equation given by Eq.\;\ref{eq:psid} for the simple Debye  process, and for the case of $\alpha=1$, $\beta\neq 1$, one recovers the deformed model reported in \cite{ieeeted2} by some of us, which admits a solution in terms of the $q$-exponential function \cite{borges1998q, borges2004possible}.  
   
 \begin{figure}[h]
\begin{center}
\includegraphics[width=0.25\textwidth,angle=-90]{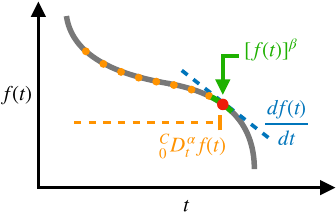}
\caption{Graphical visualization of $df(t)/dt$ vs ${}^C_0 D_t^{\alpha} f(t)$ of a function $f(t)$ at a given point of interest as marked in the figure:  (i) $df(t)/dt$ is the first-order, local derivative  of  $f(t)$ which is equal to the slope of the tangent line at that point and depends on the values of the function at $t$ and $t-dt$, while (ii) ${}^C_0 D_t^{\alpha} f(t)$ is the fractional-order, non-local derivative which invokes not just the immediate past of the function proceeding the instant $t$ but  the whole past history of the function. $[f(t)]^{\beta}$ is represented by a curved portion of $f(t)$.}
\label{fig1}
\end{center}
\end{figure}  
   
In order to illustrate schematically these functional operations,  Fig.\;\ref{fig1} shows a graph of a continuous function $f(t)$ in gray, and a specific point of interest ($t,f(t)$) marked in red. The slope of the tangent line in blue is equal to the (local) first-order derivative of the function  at that  point. This derivative requires the knowledge of $f(t)$ at the instance of time $t$ and another at $t-dt$.   However, the fractional derivative of order $\alpha$ corresponds to a weighted sum of the first-order derivative of $f(t)$ with the power law kernel $k(t,\tau)= (t-\tau)^{-\alpha}/\Gamma(1-\alpha)$ from $t=0$ till the current instance of time \cite{allagui2021possibility, memQ, memoryAPL}. The fractional derivative  is non-local and depends not only on the immediate past of the function but on its all prior history  up to the point of interest (orange representation in Fig.\;\ref{fig1}). We also plot on the same graph a green curved line around the marked point representing $[f(t)]^{\beta}$.  Eq.\;\ref{eqab} relates  the non-local fractional time derivative of $f(t)$ with  itself elevated to another arbitrary  fractional power.

\begin{figure}[h]
\begin{center}
\includegraphics[width=0.35\textwidth]{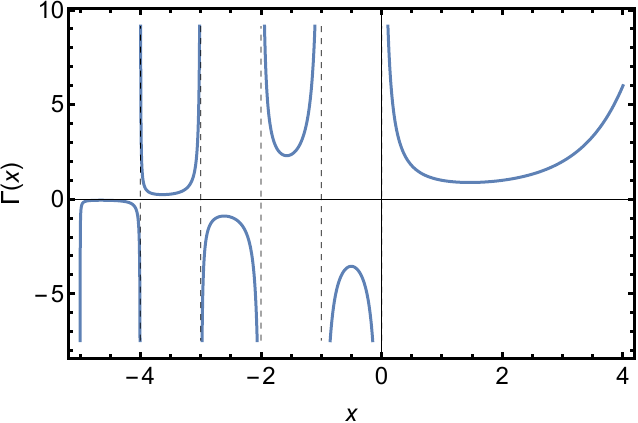}
\caption{Plot of $\Gamma(x)$ vs. real $x$ in the interval -5 to +4}
\label{fig2}
\end{center}
\end{figure}

 Now for the explicit solution of Eq.\;\ref{eqab}, we  follow the procedure of Tarasov  \cite{tarasov2020exact},
where we seek a solution in the power law form:
\begin{equation}
    \Psi(t) = C (t-t_0)^p
    \label{eq:pl0}
\end{equation}
where $t>t_0$, and $p, C \in \mathbb{R}$ ($C>0$). For simplicity, we set $t_0=0$. 
We  use  the Caputo fractional derivative of a  power law which gives\;\cite{diethelm2002analysis}:
\begin{equation}
^C_0D_{t}^{\alpha} (t^p) = 
     \frac{\Gamma(1+p)}{\Gamma(1+p-\alpha)} t^{p-\alpha}       
\label{WH}
\end{equation}
if (i) $p \in \mathbb{N}$, $p\geqslant m $   or if (ii)   
$p \notin \mathbb{N}, p > m-1 $, 
otherwise if  (iii)  
$  p \in \mathbb{N}$ and $p<m$ we have:
\begin{equation}
    ^C_0D_t^{\alpha} (t^p) =0
\end{equation}
For all other cases, i.e. when (iv) $p<m-1$, $p \neq 0,1,2,\ldots,m-1$, the integral in the expression of the Caputo fractional derivative is improper and divergent \cite{tarasov2020exact}. Plugging the result given by Eq.\;\ref{WH} into  Eq.\;\ref{eqab} leads to   the explicit solution of Eq.\;\ref{eqab} to be  \cite{diethelm2002analysis, kilbas2001differential,  1570854176956681216}:
\begin{equation}
    \Psi(t) = \left[ \frac{\Gamma(1+p)}{q_0\Gamma(1+p-\alpha)} \right]^{\frac{1}{\beta-1}}  t^{\frac{\alpha}{1-\beta}}
    \label{eq:pl}
\end{equation}
with the conditions that $\beta \in \mathbb{R}$ ($\beta \neq 0,1$), $q_0 \in \mathbb{R}$ ($q_0 \neq 0$) and $p = \alpha/(1-\beta)$. Furthermore, we should have for physical reasons:
\begin{equation}
C=\left[ \frac{\Gamma(1+p)}{q_0\Gamma(1+p-\alpha)} \right]^{\frac{1}{\beta-1}} > 0
\label{eq:C}
\end{equation} 
and to indicate a decaying relaxation profile, the power coefficient $p$ should be restricted to real negative values only, i.e.: 
\begin{equation}
\frac{\alpha}{1-\beta} <0
\label{eq:p}
\end{equation}
It is worth reminding that the gamma function $\Gamma(x)$ is a meromorphic function with no zeros, and with simple poles of residue $(-1)^n/n!$ at $x=-n$ ($n\in \mathbb{N}$):
\begin{equation}
\lim_{x\to -n} \frac{1}{\Gamma(x)}=0
\end{equation}
For visualization purposes, we show a plot of $\Gamma(x)$ for real values of $x$ in the interval -5 to +4 in Fig.\;\ref{fig2}, where it is clear that at negative integer points the function   is discontinuous and asymptotically approaches infinity. 
Thus, if  we take the  arguments of the two gamma functions in the ratio of Eq.\;\ref{eq:C} to be both  real positive, with the condition given in\;\ref{eq:p},  we must have for $\beta$:
\begin{equation}
\frac{1}{1-\alpha} < \beta 
\label{eqbeta}
\end{equation}
 with $0<\alpha<1$.
 If now both arguments of the gamma terms are bound between -1 and 0, which also satisfies inequality of Eq.\;\ref{eq:C}, we obtain the condition on $\beta$ to be:
\begin{equation}
\frac{2}{2-\alpha} < \beta < {1+\alpha}
\end{equation} 
 This can be generalized   to:
\begin{equation}
\frac{n}{n-\alpha} < \beta < \frac{n-1+\alpha}{n-1}
\label{eq:beta}
\end{equation}  
for  $n=1,2,\ldots$, always with $0<\alpha<1$. The conditions on the parameters $C>0$ and $p<0$ can not be fulfilled  if we choose  two different intervals for the arguments of the gamma functions in Eq.\;\ref{eq:C} while ensuring that they are both of the same sign, unless $\alpha$ is allowed to  not be restricted to values between zero and one.
 
\begin{figure}[t]
\begin{center}
\includegraphics[width=0.475\textwidth]{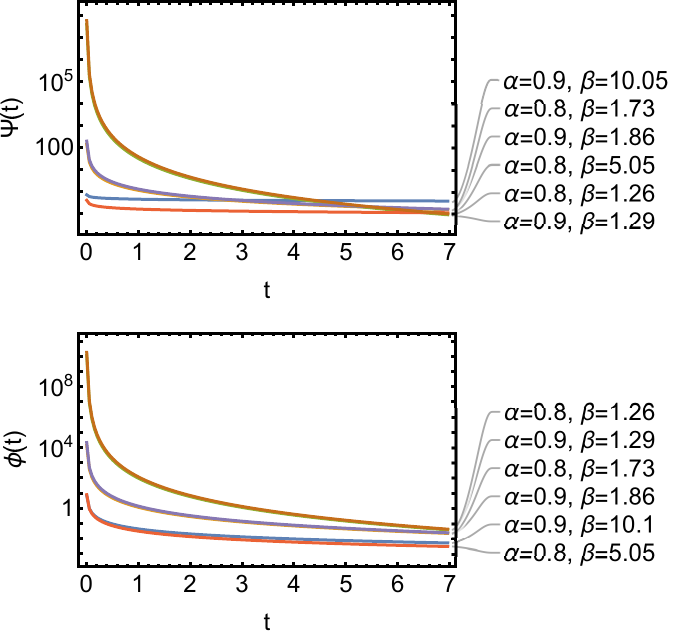}
\caption{Simulation plots of the polarization function $\Psi(t)$ and $\phi(t)$ given by Eq.\;\ref{eq:pl} and 39 vs. $t$ for different combinations of values for the parameters $\alpha$ and $\beta$.}
\label{fig3}
\end{center}
\end{figure}

\begin{figure}[!h]
\begin{center}
\includegraphics[width=0.475\textwidth]{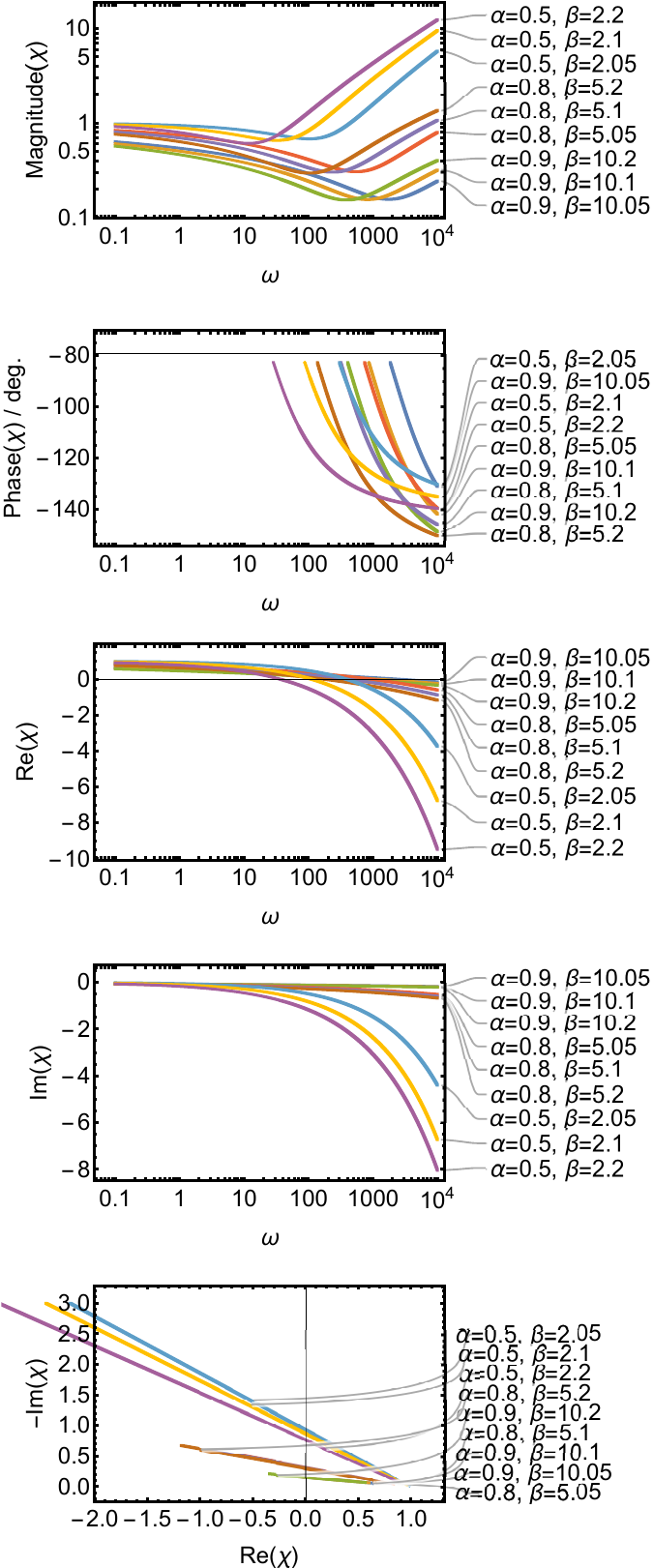}
\caption{Simulation plots of the complex susceptibility function $\tilde{\chi}(s)$ ($s=j \omega$) given by Eq. 38 in terms of Bode plots of (a) magnitude, (b) phase, (c) real and (d) imaginary part or (e) Nyquist diagram of imaginary part vs. real part, for different combinations of values for the parameters $\alpha$ and $\beta$}
\label{fig4}
\end{center}
\end{figure}

 Now using   Eqs,\;\ref{eq3} and\;\ref{eq1} giving $\tilde{\chi}(s)$ as:
 \begin{equation}
\tilde{\chi}(s) = 1 -s  \mathcal{L}  \left[  {\Psi}(t) ; s \right]
\end{equation}
 the corresponding susceptibility function associated with the polarization function $ \Psi(t)$ in Eq.\;\ref{eq:pl} is found    to be:
 \begin{equation}
\tilde{\chi}(s) =1 - \Gamma\left[
  1 + \frac{\alpha}{1-\beta}\right] 
  \left(
  \frac{\Gamma\left[1 + \frac{\alpha}{1 - \beta} \right]}
  {q_0 \Gamma\left[\frac{\beta-1  - \alpha \beta}{\beta-1 } \right] }
  \right)^{\frac{1}{\beta -1}}
  s^{\frac{\alpha}{\beta-1 }} 
  \label{eq:xi}
\end{equation}
with the same conditions outlined above on the parameters $\alpha$ and $q_0$.
 However, knowing that the LT of  a power law function of the form $f(t)=t^{\nu}$ is $\Gamma(1+\nu) s^{-\nu-1}$ valid only for $\text{Re}(\nu)>-1$, this makes the condition on $\beta$ to be $\beta>1/(1-\alpha)$ (Eq.\;\ref{eqbeta}) only.

Finally, for the sake of completeness, it is straightforward to obtain by simple derivation of the first order (Eq.\;\ref{eq3})  the pulse-response function of the system as:
\begin{equation}
    \phi(t) = \frac{\alpha}{\beta-1} \left[ \frac{\Gamma\left(1+\frac{\alpha}{1-\beta}\right)}{q_0\Gamma\left(\frac{\beta-1-\alpha \beta}{\beta-1} \right)} \right]^{\frac{1}{\beta-1}}   t^{\frac{\alpha}{1-\beta}-1}
    \label{eq:phit}
\end{equation}

\section{Discussion}

Fig.\;\ref{fig3} shows simulation plots of the polarization function $ \Psi(t)$ (Eq.\;\ref{eq:pl})  and the response function $\phi(t)$ (Eq.\;\ref{eq:phit}), both vs. time $t$,  and  
for different combinations of values for the parameters $\alpha$ and $\beta$. The values of $\alpha$ are chosen to be 0.9, 0.8, and 0.5 whereas the values we assigned to  $\beta$ are obtained according to the inequalities in Eq.\;\ref{eq:beta}  with different values of $n$ (1, 2 and 4). The value of $q_0$ is set to one in all plots. 
  The   power-law profiles of $ \Psi(t)$, parameterized with $\alpha$ and $\beta$, can be viewed as the generalization of the traditional exponential decay of Debye  by spreading out the function over time, as does  the Cole-Cole function as a matter of fact. We recall that the leading asymptotics of the Mittag-Leffler function in Eq.\;\ref{eq:Psialpha} for the Cole-Cole response are \cite{mainardi1996fractional}: 
  \begin{equation}
  E_{\alpha,1}^1\left[ -(t/\tau_{\alpha})^{\alpha} \right] \approx
 1-\frac{(t/\tau_{\alpha})^{\alpha}}{ \Gamma(1+\alpha)},\;\; t/\tau_{\alpha} \to 0
\end{equation}
and
  \begin{equation}
  E_{\alpha,1}^1\left[ -(t/\tau_{\alpha})^{\alpha} \right] \approx
 \frac{(t/\tau_{\alpha})^{-\alpha}}{ \Gamma(1-\alpha)},\;\; t/\tau_{\alpha} \to \infty
\end{equation}
evidenced in experimental responses of capacitive systems of very different nature \cite{balaguera2024limit,balaguera2024time}. However, our model offers an extra degree of freedom when compared to the latter, which appears in both the exponent and the scaling factor   of $ \Psi(t)$. The plots of $\phi(t)$ provide a complementary assessment of the system to those provided by $ \Psi(t)$,  as they indicate a  high rates of decay at smaller times compared to the much smaller decays at larger times, which is different  from  the exponential function behavior. In effect, the value of $n$ to constrain the selection of $\beta$ has a large impact on both the scale of variation of $ \Psi(t)$ and $\phi(t)$.

The effect of the extra degree-of-freedom provided by the  parameter $\beta$ is best visualized and  interpreted from the frequency-domain plots. From Fig.\;\ref{fig4}, which depicts the complex susceptibility function $\tilde{\chi}(s)$ ($s=j \omega$) given by Eq.\;\ref{eq:xi} for the same combinations of ($\alpha,\beta$) values above, we see first that at close to zero frequency, all curves converge towards to a resistance (from an electrical point of view) of one as it should be. Otherwise, the derived model can lead to  negative resistance effects and capacitive behaviors with constant phase, independent of frequency, over close to the whole range of frequency covered. In particular, we see from Fig.\;\ref{fig4}(e) that the  versatility of the model allows ones to  cover these anomalous behaviors pertinent to the third quadrant of the complex plane \cite{bisquert2022hopf,vivier2022impedance}) which is not possible with the standard Cole-Cole model or the constant phase element, as well as with the theory of Davidson-Cole and Havriliak-Negami and the subsequent standard/modified Gerischer impedances, without entering negative non-physical values. We also see from the magnitude and phase plots how the value of $\beta$ for a constant value of $\alpha$ can affect the nature of the response. Finally, as we mentioned above for the case of the time-domain functions, there is an implicit relation between the scaling factor and the fractional exponent that surely needs further investigations on real spectral data.

%
%

\section{Conclusion}

In this work we derived and studied  the polarization function $ \Psi(t)$ given by Eq.\;\ref{eq:pl}, which is the explicit solution of the fractional nonlinear differential equation given by Eq.\;\ref{eqab}. This dynamical model with the two parameters $\alpha$ and $\beta$, is considered to be a possible  generalization of  the kinetic equation describing the Cole-Cole type of response, which itself is a generalization of the Debye response. The presented approach is different from the Cole-Cole, Davidson-Cole, or the Havriliak-Negami generalizations of the Debye model, which are  done by introducing fractional  power parameters to the frequency function, and not to the evolution equations. From $ \Psi(t)$ we derived the complex susceptibility function $\tilde{\chi}(s)$ (Eq.\;\ref{eq:xi}) and analyzed its magnitude, phase and Nyquist plots for different combinations of values for the parameters $\alpha$ and $\beta$. It is shown that the proposed model can act as versatile element capable of capturing behaviors in the third quadrant of the complex plane by adjusting the parameters $\alpha$ and $\beta$. This generalized model, when compared to the Cole-Cole, Davidson-Cole or Havriliak-Negami relaxation processes and the subsequent constant phase element or standard/modified Gerischer impedances, can be of great utility in fitting experimental spectra of ac conductivity and dielectric permittivity, (bio)impedance spectroscopy, viscoelasticity, and  related fields. Nonetheless, as it is still the case for the other empirical models generalizing the Debye process, the physical interpretation of the parameters $\alpha$ and $\beta$ is open for debate.

\section*{Declarations}

The authors declare that they have no known competing financial interests or personal relationships that could have appeared to influence the work reported in this paper.

 
\bibliographystyle{elsarticle-num} 



\end{document}